\documentclass[aps,pre,preprint,groupedaddress]{revtex4}

\usepackage{graphicx}
\usepackage{latexsym}
\usepackage{amsmath,amssymb}

\begin{document}


\title{
Finite-size-scaling analysis of 
the $XY$ universality class
between two and three dimensions:
An application of 
Novotny's transfer-matrix method
}

\author{Yoshihiro Nishiyama}
\affiliation{Department of Physics, Faculty of Science,
Okayama University, Okayama 700-8530, Japan.}

\date{\today}

\begin{abstract}
Based on Novotny's transfer-matrix method,
we simulated
the (stacked) triangular Ising
antiferromagnet embedded in the space
with the dimensions variable in the range $2 \le d \le 3$.
Our aim is to investigate the criticality of the
$XY$ universality class for $2 \le d \le 3$.
For that purpose, we employed 
an extended version of the
finite-size-scaling analysis developed by Novotny,
who utilized this scheme
to survey the Ising criticality (ferromagnet) for $1 \le d \le 3$.
Diagonalizing the transfer matrix for the system sizes $N$ up to
$N=17$,
we calculated the $d$-dependent correlation-length critical exponent
$\nu(d)$.
Our simulation result $\nu(d)$ appears to interpolate smoothly
the known two limiting cases, namely, the KT and $d=3$ $XY$ universality
classes,
and the intermediate behavior bears close resemblance to 
that of the analytical formula via the $1/N$-expansion technique.
Methodological details including the modifications specific
to the present model are reported.
\end{abstract}


\maketitle

\section{\label{section1}Introduction}

In analytical approaches, the spatial dimension $d$ 
is treated as 
a continuously variable parameter, and correspondingly, 
various quantities such as the critical indexes
are expressed explicitly in terms of the parameter $d$.
Such an approach allows us to see
how the criticality changes from
the classical (meanfield like) one 
as the spatial dimension deviates
from an
either lower or upper critical dimension gradually.
However, it is not quite obvious that 
such an analytical formula
could be justified (realized) by actual 
first-principles simulations 
(and hopefully by experiments) with respect to realistic lattice models.
In fact, in conventional computer-simulation approaches, one has to fix
the (embedding) spatial dimension to a certain integral value,
and thus the analysis on criticality has been restricted
to the integral values of $d$ inevitably.

An attempt to circumvent such a restriction was
made by Novotny \cite{Novotny90,Novotny92,Novotny93a,Novotny93b}.
His approach stems on a very formal expression for the
transfer matrix so that the embedding spatial dimension
can be varied continuously.
(We explain his method in the next section.
As anticipated naturally, this method is also of use 
in studying high dimensional ($d \ge 3$) systems.
We refer readers to the references \cite{Novotny90,Nishiyama04} 
for this development.)
Based on this formulation, he performed an extensive computer simulation,
and surveyed the 
criticality of the Ising ferromagnet for $1 \le d \le 3$.
Astonishingly enough, he found
that the numerical result is well described by both the
$1+\epsilon$ and $4-\epsilon$ expansion formulas.
In other words, his result clarifies that the analytical
formulas for fractional values of $d$ are meaningful in the sense that
they are reproduced by the first-principles-simulation scheme.
(Strictly speaking, he utilized two distinctive approaches
to control the embedding spatial dimension.
In Ref. \cite{Novotny92}, he varies the ``connectivity'' of the lattice,
whereas in Refs. \cite{Novotny93a,Novotny93b}, he twists
the boundary condition to control the 
magnetic-domain-wall undulations.)
Here, we stress that the Novotny approach is
not a mere dimensional interpolation (crossover) that has been 
studied extensively in the past study \cite{Landau95,Ruge95}.

In this paper, we apply Novotny's method to the ``stacked''
triangular Ising antiferromagnet 
\cite{Landau83,Kitatani83,Miyashita91,Queiroz95,Miyashita97,%
Blankschtein84,Heinonen89,Yamada90,Netz91,Bunker93,Gaulin95,%
Plumer93,Plumer95,Dantchev04}
embedded in the space with the variable dimensions $2 \le d \le 3$.
Because of its $Z_6$ invariance, the model should
exhibit the $XY$ universality class at the magnetic transition point
\cite{Jose77,Tobochnik82,Rujan81,Challa86,Irakliotis93}.
Our aim is to examine whether his method is applicable to
generic problems other than the Ising universality class.
We calculate the $d$-dependent correlation-length critical exponent.
Thereby, we will show that
the simulation result is comparable with the analytical $1/N$-expansion result
up to O$(1/N)$ \cite{Ma72,Abe72,Suzuki72,Ferrel72}.

In fairness, it has to be mentioned that
the critical phenomena
for non-integer dimensions were studied
extensively in the past \cite{Gefen84,Lin86,Taguchi87,Costa87,Hao97}.
In these works, the authors set up their lattice models
on a fractal structure (the Sierpinski gasket) in order to
realize a magnetism in the fractional spatial dimensions.
Here, we stress that in our approach,
the spatial dimension can be varied
continuously within a range.

The rest of this paper is organized as follows.
In the next section, we explain how we constructed the
transfer matrix for the stacked triangular antiferromagnet
in $2 \le d \le 3$.
In Sec. \ref{section3},
we present the numerical results.
Managing an extended finite-size scaling analysis,
we estimate the correlation-length critical exponent 
in the range $2 \le d \le 3$.
In the last section, we present summary and discussions.

\section{\label{section2}
Construction of the transfer matrix
for the ``stacked'' triangular antiferromagnet
in $2 \le d \le 3$}

In this section, we set up the transfer matrix formalism
to simulate the ``stacked'' triangular antiferromagnet 
embedded in the dimensions $2\le d \le 3$.
Our formalism is based on Novotny's idea \cite{Novotny90}, with which
he studied the Ising ferromagnet on (hyper) cubic lattices;
in our preceding paper \cite{Nishiyama04}, we adopted his idea to the 
$d=3$ Ising ferromagnet
with plaquette-type interactions.
Before going into details, we first set up a basis of our scheme;
namely, we dwell on the particular $d=3$ case.
Then, we extend this preliminary basis to
incorporate the embedding-spatial-dimension variation.
We will also 
provide a number of technical modifications to improve
the efficiency of the numerical simulation.

\begin{figure}
\includegraphics[width=70mm]{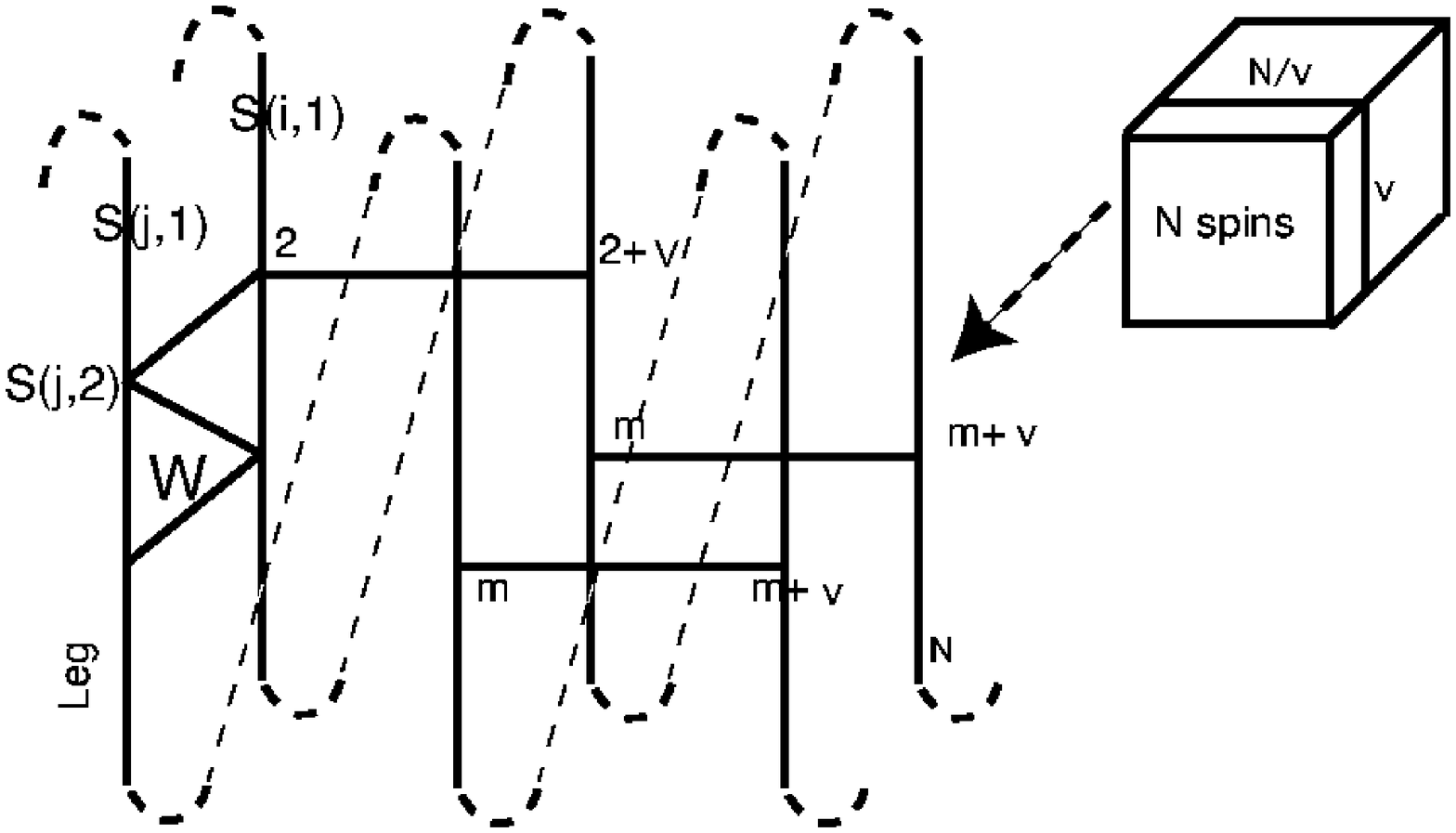}%
\caption{\label{figure1}
A schematic drawing of the construction
of the transfer matrix 
for the ``stacked'' triangular antiferromagnet.
Our scheme is based on Novotny's idea  \cite{Novotny90}.
The sheet of triangular antiferromagnet extends
along the ``leg.''
The leg is folded into a rectangular shape:
To be specific,
with use of the translation operator $P^v$
($v$: interaction distance), we build a bridge between the 
$v$th neighbor spins along the leg.
}
\end{figure}

We decompose the transfer matrix into the following two contributions,
\begin{equation}
T = T^\perp \odot T^\parallel ,
\end{equation}
where the symbol $\odot$ denotes the Hadamard (element by element) matrix
multiplication;
note that the multiplication of local Boltzmann factors yields
the global Boltzmann weight.
As explained below (see also Fig. \ref{figure1} for the geometrical
structure of our finite-size cluster),
the decomposed parts $T^\perp$ and $T^\parallel$ account for the
contributions from the
intra and inter plane (triangular lattice) interactions, respectively.

First, let us consider the component $T^\parallel$.
The matrix elements are given by the formula,
\begin{equation}
T^\parallel_{ij}=
   \langle i|A|j \rangle=
   W^{S(j,1)S(j,2)}_{S(i,1)S(i,2)}
   W^{S(j,2)S(j,3)}_{S(i,2)S(i,3)}  \cdots
   W^{S(j,N)S(j,1)}_{S(i,N)S(i,1)}
,
\end{equation}
where the indexes $i$ and $j$ specify the Ising spin configuration for 
both sides of the transfer-matrix slice;
see Fig. \ref{figure1}.
More specifically, we consider $N$ spins for the transfer-matrix slice,
and the index $i$ denotes a spin configuration 
$\{S(i,1), \dots , S(i,N) \}$
arranged along the ``leg.''
The factor $W^{S_3 S_4}_{S_1 S_2}$ 
stands for the local Boltzmann weight for a unit cell
of the triangular lattice with the corner spins 
$\{ S_1,S_2,S_3,S_4 \}$.
Explicitly, it is given by the following form,
\begin{equation}
W^{S_3 S_4}_{S_1 S_2} =
  \exp \left[
  -\frac{1}{T}
           \left(
 \frac{J}{2} (S_1 S_2 +S_2 S_4+S_4 S_3 + S_3 S_1)
   +J S_1 S_4
           \right)
       \right]     .
\end{equation}
(The denominator of the coupling constant is intended to avoid double 
counting.)
Here, the parameter $T$ denotes the temperature,
and the parameter $J$ stands for the intra-plane antiferromagnetic
interaction constant.
Hereafter,
we choose $J$ as the unit of energy; namely, we set $J=1$.
It is to be noted that the component
$T^\parallel$ (with $T^\perp$ ignored) leads to the transfer-matrix 
for a sheet of triangular
antiferromagnet.
In other words,
the remaining component $T^\perp$ should raise the dimensionality
to $d=3$ through introducing the inter-plane interactions.
This is an essential idea of the Novotny method \cite{Novotny92}.

Second, we consider the component $T^\perp$ that
accounts for the inter-plane interaction.
The explicit matrix elements are
given by the following formula,
\begin{equation}
T^\perp_{ij}
  =  \langle i| B P^v |i \rangle 
  ,
\end{equation}
with the interaction distance $v$.
The matrix $B$ is given by the formula,
\begin{equation}
\langle i | B |j \rangle = 
   W_{S(i,1)S(j,1)}^\perp
   W_{S(i,2)S(j,2)}^\perp  \cdots
   W_{S(i,3)S(j,3)}^\perp   ,
\end{equation}
with
$W_{S_1 S_2}^\perp=\exp(-j S_1 S_2/T)$ and
the inter-plane interaction $j$.
The matrix $P$ denotes the translational operator:
That is, with one operation of $P$, a spin arrangement 
$\{ S(i,m) \}$ shifts to 
$\{ S(i,m+1) \}$;
note that the periodic boundary condition is imposed.
An explicit representation of $P$ is given in our preceding paper
\cite{Nishiyama04}.
Because of the insertion of $P^v$,
the interaction $B$ bridges the $v$th neighbor pairs along the leg, 
and so
it brings about the desired inter-plane interactions.
As a matter of fact, in Fig. \ref{figure1},
we notice that the alignment of spins is folded into a 
rectangular shape with the edge lengths $v$ and $N/v$.
It is an essential idea of Novotny that
the operation $P^v$ is still meaningful,
although the power $v$ is not an integral value.
This rather remarkable fact renders freedom that one can 
construct the $d=3$ transfer-matrix systematically
with arbitrary number of spins $N$.

Based on the above formalism, we readily simulate the stacked triangular 
antiferromagnet in $d=3$.
In the following,
we propose a scheme to tune the embedding spatial 
dimension continuously.
Moreover, we will also provide a number of technical modifications,
aiming to improve the efficiency of the simulation.

There are two controllable parameters for the dimension variation.
That is,
the inter-plane interaction $j$ and the interaction distance $v$:
Apparently, the limit $j \to 0$
reduces the system to a sheet of triangular lattice.
On the other hand,
for large $v$, the stack width $N/v$ decreases,
and eventually, at $v=N$, the system reduces to a sheet of triangular
lattice as well; note the identity $P^N=1$ owing to the periodic boundary condition.
In this paper, we adopt the former scheme.
Namely,
we will tune the parameter $j$, fixing the interaction distance $v$
to a moderate value $v=0.27N$.
This choice is based on our observation that the finite-size-scaling
behaviors become quite systematic for $v \approx N/n$ ($n$: integer),
particularly at $n=4$.

Lastly, let us explain a number of technical modifications to improve the
efficiency of the simulation:
We propose the following replacement,
\begin{equation}
T^\perp_{ij} = \langle i|B P^v|i\rangle 
  \to  T^{\perp}_{ij} (v)=
 \langle i|B P^v|i\rangle  \langle i| P^{-v} B |i\rangle   .
\end{equation}
(Note that correspondingly, we need to replace the temperature $T$ with $2T$
in order to compensate the duplication.)
With this trick, the transfer-matrix elements become real:
Otherwise, the elements are complex for even values of $N$.
As a matter of fact, in the past simulations \cite{Novotny90,Novotny92,Nishiyama04},
those cases of
even values of $N$ were excluded.
Such an exclusion is obviously disadvantageous in the subsequent data analysis, 
because the available systems sizes are restricted severely.
In our simulation, because of the above trick,
we are able to consider arbitrary system sizes.
In addition to this, we symmetrize the transfer matrix 
\cite{Novotny92} with the following replacement,
\begin{equation}
T^{\perp}(v) \to T^{\perp} (v) \odot T^{\perp}(-v)   .
\end{equation}
(Similarly as the above, we need to re-define the temperature $T\to 2T$.)
With this symmetrization, the symmetry of the descending 
($j=N,N-1,\dots$) and ascending ($j=1,2,\dots$) directions along the leg
become completely restored.
We observed a significant improvement of 
the finite-size scaling behavior due to this symmetrization.


\section{\label{section3}Numerical results}

In the preceding section, we developed a transfer-matrix formalism
for the stacked triangular antiferromagnet embedded in 
the fractional spatial dimensions $2 \le d \le 3$.
In this section, we present the numerical results
calculated by means of 
the exact-diagonalization method
for the system sizes up to $N=17$.
We analyze the data with the extended finite-size scaling analysis \cite{Novotny92}
which allows us to 
estimate the ``effective'' dimension $d_{eff}$.
The effective dimension plays a significant role
in the subsequent analysis of
the criticality of the magnetic transition.

\subsection{Effective dimension: 
Extended finite-size scaling analysis \cite{Novotny92}}

Before going into detailed analysis on the criticality,
we need to estimate the effective dimension $d_{eff}$ \cite{Novotny92}:
At the critical point, 
the correlation length should be comparable to
the linear dimension of the finite cluster.
Hence, the correlation length $\xi$ should obey the 
formula $\xi \sim N^{1/(d_{eff}-1)}$;
note that a transfer-matrix slice contains $N$ lattice points,
and its embedding spatial dimension  
should be $d_{eff}-1$.
This formula immediately yields an estimate for the
effective dimension,
\begin{equation}
d_{eff}^{N,N'}(T)= 
      \frac{1}{\ln (\xi_N(T)/\xi_{N'}(T))/ \ln(N/N')}  +1 ,
\end{equation}
for a pair of system sizes $(N,N')$.
By means of the transfer-matrix method,
the correlation length is calculated immediately:
Using the largest and the next-largest eigenvalues,
namely, $\lambda_1$ and $\lambda_2$ of the transfer matrix, 
we obtain the correlation length as
$\xi=1/\ln(\lambda_1/\lambda_2)$.
Provided by this,
according to Ref. \cite{Novotny92},
we are able to determine both the critical temperature
$T_c(N_1,N_2;N_3,N_4)$
and the effective dimension $d_{eff}(N_1,N_2;N_3,N_4)$ so that
they satisfy the following equation,
\begin{equation}
d_{eff}(N_1,N_2;N_3,N_4) = 
d_{eff}^{N_1,N_2}(T_c(N_1,N_2;N_3,N_4)) =
d_{eff}^{N_3,N_4}(T_c(N_1,N_2;N_3,N_4)) ,
\end{equation}
for the set of the system sizes $(N_1,N_2;N_3,N_4)$.
To summarize, in the extended finite-size scaling analysis, 
the spatial dimension is not a given constant, but a parameter
that is to be determined {\it a posteriori}
with the data analysis of the correlation length $\xi$.
As mentioned in the above, in the transfer-matrix method,
the correlation length is calculated quite straightforwardly.
In that sense, the transfer-matrix approach is suitable to this type 
of finite-size-scaling analysis.

\begin{figure}
\includegraphics[width=70mm]{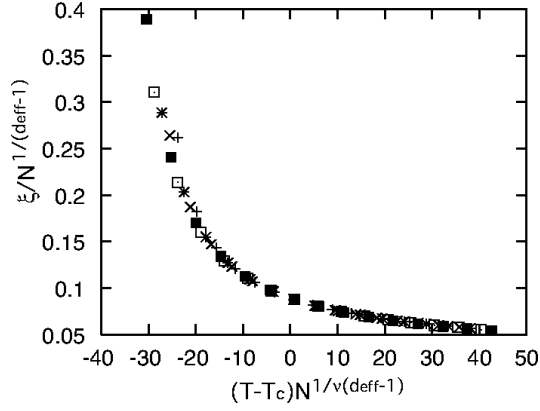}%
\caption{\label{figure2}
Scaling plot of the correlation length 
$(T-T_c)N^{1/\nu(d_{eff}-1)}$-$\xi/N^{1/(d_{eff}-1)}$ is shown
for $j=1.5$ and the system sizes $N=13,14,\dots,17$.
The symbols $+$, $\times$, $*$, $\Box$, and $\blacksquare$
denote the system sizes of $N=13$, $14$, $15$, $16$, and $17$,
respectively.
The scaling parameters are set to be $1/\nu=1.36$, $d_{eff}=2.50$,
and $T_c=5.13$; see text for details.
We see that a good data collapse is achieved under a fractional value
of the effective dimension $d_{eff}=2.50$.
}
\end{figure}

In order to examine the validity of the scaling parameters,
$d_{eff}$ and $T_c$, determined with the above method,
we plotted,
in Fig. \ref{figure2},
the scaled correlation length 
$(T-T_c)N^{1/(\nu(d_{eff}-1))}$-$\xi /N^{1/(d_{eff}-1)}$
for $d_{eff}=2.50$, $T_c=5.13$, $1/\nu=1.36$, $j=1.5$,
and $N=13,14,\dots,17$.
These scaling parameters, namely, $d_{eff}$ and $T_c$, are determined 
from the set of system
sizes $(14,16;13,16)$ via the extended finite-size-scaling analysis.
(We explain how we determined $1/\nu=1.36$ afterward.)
We see that the scaled data collapse into a scaling function quite satisfactorily.
Hence, we confirm that 
the scaling parameters, $d_{eff}=2.50$  and $T_c=5.13$, are indeed meaningful.
More significantly, we stress that our simulation data 
should be described under the assumption that the effective dimension
takes such a fractional value.

\begin{figure}
\includegraphics[width=70mm]{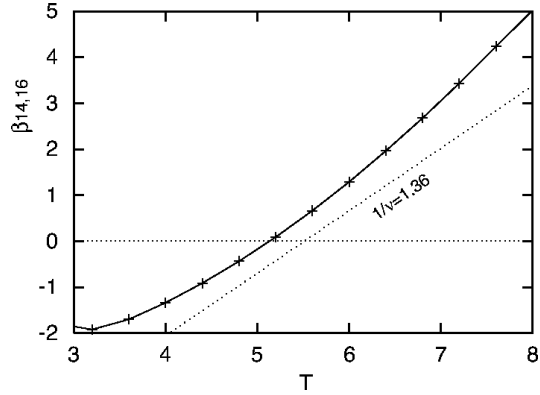}%
\caption{\label{figure3}
We plotted the beta function, $\beta_{14,16}(T)$ (\ref{beta_function}),
for the same parameters as Fig. \ref{figure2}.
From the slope at the transition point, 
we estimate the correlation-length critical exponent
as $1/\nu=1.36$.
}
\end{figure}

In order to analyze the criticality further in detail,
we calculated the Roomany-Wyld approximative $\beta$ function,
which is given by the following formula
\cite{Roomany80},
\begin{equation}
\beta_{N,N'}(T)  =
       \frac{1-(d_{eff}-1)\ln(\xi_N/\xi_{N'})/\ln(N/N')}
         {\sqrt{\partial_T \xi_N(T) \partial_T \xi_{N'}(T)
              /  \xi_N(T)/  \xi_{N'}(T)}}  .
\label{beta_function}
\end{equation}
In Fig. \ref{figure3}, we plotted $\beta_{14,16}(T)$ for 
the same parameters as those of Fig. \ref{figure2}.
The functional form of this $\beta$ function seems to be almost straight,
indicating that the corrections to the finite-size scaling are almost negligible.
This fact accounts for the good data collapse of Fig. \ref{figure2} shown above.
From the slope of this $\beta$ function
at the transition point $T=T_c$, we are able to estimate the 
inverse of the correlation-length critical exponent as $1/\nu_{14,16}=1.36$.
To summarize, we estimated the exponent $1/\nu_{N,N'}$
from the set of system sizes $(N,N')$.
In prior to this analysis, we should determine
the scaling parameters, $d_{eff}$ and $T_c$,
from the extended finite size scaling analysis for
$(N_1,N_2;N_3,N_4)$.
We will exploit the $d_{eff}$-dependence of $1/\nu$ in the 
next subsection.

\begin{figure}
\includegraphics[width=70mm]{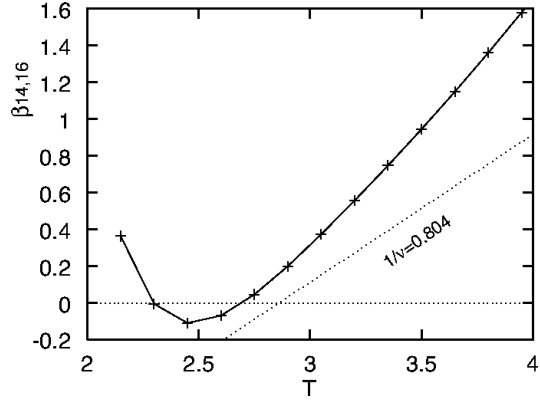}%
\caption{\label{figure4}
The beta function, $\beta_{14,16}(T)$ (\ref{beta_function}), is plotted for 
$j=0.9$ and $d_{eff}=2.07$.
We notice that for $d_{eff} \approx 2$,
the beta function gets curved, indicating that a
nonstandard (essentially singular) criticality emerges.
From the slope at the transition point, 
we estimate the correlation-length critical exponent
as $1/\nu=0.804$.
}
\end{figure}

Lastly, we exploit the region in close vicinity of the lower
critical dimension $d=2$, at which the KT-type singularity
should emerge.
In Fig. \ref{figure4}, we present the function $\beta_{14,16}(T)$
for $j=0.9$, $T_c=2.74$, and $d_{eff}=2.07$;
these scaling parameters were determined from the set of system
sizes $(14,16;15,17)$.
In contrast to the behavior shown in Fig. \ref{figure3},
the $\beta$ function is curved particularly in the vicinity 
of the transition point.
Actually, we estimate the slope (critical exponent)
as $1 / \nu_{14,16}=0.804$ which is considerably suppressed,
compared with that of Fig. \ref{figure3}.
This feature indicates that an 
essentially singular-type critical behavior emerges as $d_{eff}\to 2$.
For $T<T_c$, the $\beta$ function starts to increase, 
and eventually,
it becomes even positive in the low-temperature regime.
Such a feature may reflect an instability to the 
$Z_6$-symmetry-broken phase.
Actually, it has been known \cite{Jose77} that right at $d=2$, 
an additional phase transition of the KT type takes place
at a low temperature, where the $Z_6$-symmetry-breaking field
becomes marginally relevant.
The simulation data around this regime may be affected by the 
notorious logarithmic corrections to the finite-size-scaling
behavior, that are inherent to the KT-type critical phenomenon.

\subsection{Correlation-length critical exponent $1/\nu(d_{eff})$ for $d_{eff}$}

In the above, we analyzed the criticality of the magnetic transition
for $j=1.5$ and $0.9$ in terms of the effective dimension $d_{eff}$.
Managing the similar analysis for various $j$,
we are able to survey the $d_{eff}$-dependence of the critical exponent $1/\nu(d_{eff})$.
In Fig. \ref{figure5}, we plotted 
the critical exponent $1/\nu$ for various $d_{eff}$.
The exponent was determined from the set of system sizes,
($\blacksquare$) $(14,16)$, ($\times$) $(15,17)$, ($*$) $(14,16)$, and ($\Box$) $(13,17)$,
for respective symbols, and
the corresponding scaling parameters, namely,
$d_{eff}$ and $T_c$, had been determined from the set of system sizes
($\blacksquare$) $(14,16;15,17)$, 
($\times$) $(15,17;13,17)$,
($*$) $(14,16;13,16)$, and
($\Box$) $(15,17;13,17)$, respectively.
We also plotted a result $1/\nu=1.48909(60)$ 
for the $d=3$ $XY$ universality class \cite{Compostrini01}
with the symbol $+$.
We notice that our numerical results interpolate smoothly the known limiting
cases of KT ($1/\nu=0$) and $d=3$ $XY$ universality classes.
As for a comparison, with a dotted line, we presented the $1/N$-expansion-approximation result
up to O$(1/N)$ \cite{Ma72,Abe72,Suzuki72,Ferrel72},
\begin{equation}
\frac{1}{\nu}
   = d-2 + \frac{2(3-\epsilon)(2-\epsilon)}{N(4-\epsilon)}
    \frac{4 \sin \frac{\pi\epsilon}{2}\Gamma(2-\epsilon)}
             {\pi\Gamma(1-\frac{\epsilon}{2})\Gamma(2-\frac{\epsilon}{2})}
  ,
\end{equation}
with $N=2$ and $\epsilon=d-2$.
Rather remarkably, we see that the simulation data and the 
$1/N$-expansion result exhibit similar intermediate behaviors.
In other words, we can make contact with such a dimensional-regularized
analytical expression
via the computer simulation calculation.
On closer inspection, however, it seems that our first-principles 
simulation predicts even more convex-like functional form.
Actually, our simulation result suggests a notable steep increase around $d=2$.

It is to be noted that our data extend to the regime exceeding the 
threshold $d=3$.
As a matter of fact, 
we intended to cover the parameter range $ 2 \le d \le 3$,
when we constructed the transfer matrix in Sec. \ref{section2}.
However, such a feature was also 
observed in the preceding study of the Ising
ferromagnet \cite{Novotny93a}.
In the study, the author reported that the effective
dimension does exceed the intended range, and 
astonishingly enough, the data are 
still in good agreement
with the $(4-\epsilon)$-expansion result.
We expect that our result makes sense
even for $d_{eff}>3$ as well.
Our data seem to approach to the meanfield value $1/\nu=2$ as $d \to 4$
rather directly than the $1/N$-expansion-approximation result.

In close vicinity of the lower critical dimension,
particularly, around $d_{eff} \approx 2.3$,
the numerical data turn out to be scattered.
As noted in the preceding subsection, this data scatter
should be attributed to the logarithmic corrections to the
finite-size scaling,
which are inherent to the KT-type critical behavior.

\section{\label{section4} Summary and discussions}

\begin{figure}
\includegraphics[width=70mm]{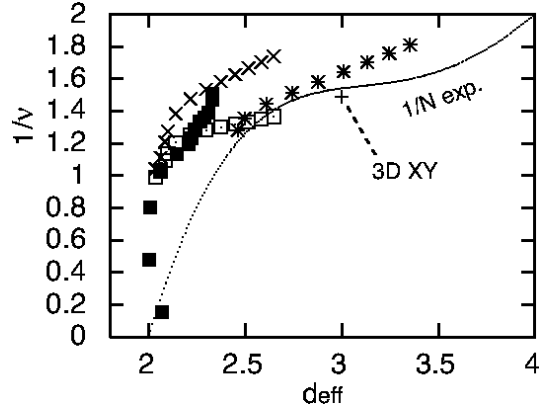}%
\caption{\label{figure5}
The inverse of the correlation-length critical exponent $1/\nu(d_{eff})$ is plotted
for the effective dimension $d_{eff}$.
The exponent is determined from the pair of system sizes,
($\blacksquare$) $(14,16)$ under the extended scaling analysis \cite{Novotny92} for $(14,16;15,17)$,
($\times$) $(15,17)$ with $(15,17;13,17)$, 
($*$) $(14,16)$ with $(14,16;13,16)$, and
($\Box$) $(13,17)$ with $(15,17;13,17)$, for respective symbols;
see text for details.
We also plotted the $1/N$-expansion-approximation result up to O$(1/N)$ 
\cite{Ma72,Abe72,Suzuki72,Ferrel72}.
The symbol $+$ denotes a result 
for the $d=3$ $XY$ model
\cite{Compostrini01}.
}
\end{figure}

In this paper, we developed a transfer-matrix scheme to simulate
the ``stacked'' triangular antiferromagnet
embedded in $2 \le d \le 3$.
Our scheme is based on Novotny's idea, which has been applied
to the Ising ferromagnet (universality class) in
$1 \le d \le 3$ successfully \cite{Novotny92,Novotny93a,Novotny93b}.
Here, we studied the $XY$ universality class in $2 \le d \le 3$;
the triangular antiferromagnet should belong to the $XY$
universality class due to the $Z_6$ symmetry
\cite{Jose77,Tobochnik82,Rujan81,Challa86,Irakliotis93}.
The numerical data are analyzed in terms of the
extended finite-size-scaling method \cite{Novotny92}, which allows us
to estimate the effective dimension $d_{eff}$.
Thereby, we obtained the $d_{eff}$-dependent
correlation-length critical exponent $1/\nu(d_{eff})$; see Fig. \ref{figure5}.
We notice that our first-principles data interpolate smoothly the
known limiting cases of both KT and $d=3$ $XY$
universality classes.
Furthermore, we found that the intermediate behavior 
bears close resemblance to that of the analytical formula via
the $1/N$-expansion technique.
In other wards, by means of the computer simulation method,
we are able to check (support) the validity of the dimensional-regularized
formulas with the $\epsilon$ and $1/N$ expansion techniques.
On closer inspection, our simulation result suggests a notable steep increase
around $d=2$.

The present study on the stacked triangular antiferromagnet
shows that the Novotny method would be
generic, and it should be
applicable to a wide variety of universality classes other than the Ising universality.
So far, the transfer-matrix approach has been restricted to the problems
in two dimensions, because it requires huge computer memory space
as the system size increases.
(Although the density-matrix-renormalization-group method 
resolves this difficulty to a considerable extent,
its extension to $d=3$ is still a current topics underway
\cite{Maeshima04}.)
With the aid of the Novotny method, 
we are able to construct the transfer matrix for
$d > 2$ quite systematically with modest (actually, arbitrary)
number of constituent spins.
Moreover, we can survey the criticality even in the fractional dimensions
by means of the extended finite-size-scaling analysis.
This opens a way to re-examine the longstanding problems in three dimensions
such as the chiral universality \cite{Kawamura98}
and the Lifshitz-type multi-critical phenomenon \cite{Diehl04}.

One may wonder 
what affects the variation of $d_{eff}$ most significantly:
Actually, there have been known
two approaches 
in order to control $d_{eff}$
in the past studies of the Ising ferromagnet.
In Ref. \cite{Novotny92}, the interaction distance $v$,
in other words, the connectivity of the finite-size
cluster, is tuned carefully.
On the other hand, 
in Refs. \cite{Novotny93a,Novotny93b}, the boundary condition is twisted 
so as to control the thermal undulations of the magnetic domain walls.
In our scheme, as explained in Sec. \ref{section2},
we fixed the lattice connectivity $v$,
and rather varied the inter-plane interaction $j$:
In this sense, we took an advantage that our system
(stacked triangular antiferromagnet) is, by nature, spatially anisotropic,
and so we are able to tune the inter-plane interaction freely.
We suspect that our case may belong to the latter category.
That is, the inter-plane interaction,
somehow, controls the domain-wall undulations effectively.
This interpretation is based on the fact that
our system is subjected to the magnetic frustration due to the
triangular antiferromagnetism, and the
domain walls should be created inevitably.
It is desirable that this mechanism would be exploited 
in the future study.

\begin{acknowledgments}
This work is supported by Grant-in-Aid for
Young Scientists
(No. 15740238) from Monbu-kagakusho, Japan.
\end{acknowledgments}


\end{document}